\documentclass[12pt,thmsa]{article}
\usepackage{amssymb}

\usepackage{float}
\usepackage{sw20aip}
\usepackage{t1enc}

\input{tcilatex}
\begin{document}

\title{The Eikonal Equation in Flat Space: Null Surfaces and Their Singularities. I }
\author{Simonetta Frittelli$^{\text{a,b}}$, Ezra T. Newman$^{\text{b}},$ \and %
Gilberto Silva-Ortigoza$^{\text{b,c}}$ \\
$^{\text{a}}${\small Physics Department, Duquesne University, Pittsburgh, PA
15282,}\\
$^{\text{b}}${\small Department of Physics and Astronomy, }\\
{\small University of Pittsburgh, }\\
{\small Pittsburgh, PA 15260.}\\
$^{\text{c}}${\small Facultad de Ciencias F\'{\i }sico Matem\'{a}ticas }\\
{\small de la Universidad Aut\'{o}noma de Puebla, }\\
{\small Apartado Postal 1152, }\\
{\small 72001 Puebla, Pue., M\'{e}xico. }}
\date{Aug.12, 1998}
\maketitle

\begin{abstract}
{\small The level surfaces of solutions to the eikonal equation define null
or characteristic surfaces. In this note we study, in Minkowski space,
properties of these surfaces. In particular we are interested both in the
singularities of these ``surfaces'' (which can in general self-intersect and
be only piece-wise smooth) and in the decomposition of the null surfaces
into a one parameter family of two-dimensional wavefronts which can also
have self-intersections and singularities.}

{\small We first review a beautiful method for constructing the general
solution to}$\ ${\small the flat-space eikonal equation; it allows for
solutions either from arbitrary Cauchy data or for time independent
(stationary) solutions of the form }$S=t-S_{0}(x,y,z).${\small \ We then
apply this method to obtain global, asymptotically spherical, null surfaces
that are associated with shearing (``bad'') two-dimensional cuts of null
infinity; the surfaces are defined from the normal rays to the cut. This is
followed by a study of the caustics and singularities of these surfaces and
those of their associated wavefronts. We then treat the same set of issues
from an alternative point of view, namely from Arnold's theory of generating
families. This treatment allows one to deal (parametrically) with the
regions of self-intersection and non-smoothness of the null surfaces,
regions which are difficult to treat otherwise. }

{\small Finally, we generalize the analysis of the singularities to the case
of families of characterisitic surfaces.}
\end{abstract}

\section{Introduction}

The study of the propagation of electromagnetic wavefronts, in the
high-frequency or geometric optics limit, is almost ubiquitous in physics;
it is a basic staple of elementary physics courses, it arises in the
practical area of optical equipment, in applied subjects, too numerable to
mention in detail, involving materials with variable refractive index , in
atmospheric and astrophysical studies. They have been a prime illustration
of V.I. Arnold's theory of Lagrangian and Legendre maps\cite{A1,A2,A3,A4}.
In a different guise, similar problems arise in catastrophe theory. In
addition to the various applications to more standard physics problems, they
also play a most fundamental role in General Relativity, e.g., the
continuous propagation of the two-dimensional wavefronts, i.e., the one
parameter family of evolving wave fronts, form null (or characteristic)
three-surfaces that are determined by the dynamics of the curved space-time
in which the wavefronts propagate\cite{HF,Hasse}. In this context it forms
the basis for the theory of gravitational lensing\cite{SEF}. The converse
statement is also true, namely that sets of null surfaces define, up to a
conformal factor, the space-time geometry itself\cite{NSF1,NSF2}. In
arbitrary space-times, the high frequency limit is completely governed by
the eikonal equation,$\qquad \qquad \qquad $%
\begin{equation}
g^{ab}\partial _{a}S_{b}S=0  \label{a}
\end{equation}

\noindent where $x^{a}=(x^{i},t)$ are local space-time coordinates$,$ $%
g^{ab}(x^{c})$ is a given space-time metric and $S=S(x^{a})$. The level
surfaces of $S,$ i.e., $S(x^{a})=constant,$ define the characteristic or
null three-surfaces (or what Arnold calls the ``big wave fronts''\cite{A2})
and the $S(x^{i},t)=constant$ for \textit{constant} $t$ define the two
dimensional ``small'' wavefront in the three dimensional space of $%
t=constant $. The vector $p^{a}=g^{ab}\partial _{b}S$ is tangent to the null
geodesics that rule the characteristic surface. Though we are basically
interested in Eq.(\ref{a}) for arbitrary space-times, here we will confine
ourselves to a study of its solutions only in flat (Minkowski) space. (A
future paper, in preparation, will generalize the present material to curved
space-times.) Eq.(\ref{a}) then becomes\qquad \qquad 
\begin{equation}
\eta ^{ab}\partial _{a}S\partial _{b}S=(\partial _{t}S)^{2}-(\partial
_{x}S)^{2}-(\partial _{y}S)^{2}-(\partial _{z}S)^{2}=0.  \label{b}
\end{equation}

The level surfaces of the solutions to Eqs.(\ref{a}) or (\ref{b}), the null
surfaces, can be viewed as being generated by the evolution of 2-dimensional
wave fronts. Specifically, a wavefront evolves by following light-rays that
are normal to it, generating the null three-surface. A smooth wavefront in
three-space, in general progresses, into a singular one, either to the past
or the future, i.e., a generic null surface in space-time has singularities.
The singular wavefronts are 2-surfaces that are continuous with existing
first derivative, but where (piece-wise) the second derivative are singular,
being either undefined or infinite. The structure of the singularities are
generically cusp ridges and swallowtails. There are unstable exceptions.

A textbook example~\cite{A2,A3} of flat space singular wavefronts (and
associated big wavefront) are from imploding triaxial ellipsoids, where an
initially ellipsoidal wavefront is evolved inwardly, it self-intersects for
some finite period of time and eventually expands out to infinity, becoming
spherical in the limit.

The singularities of wavefronts are also interpreted as the location of
focusing regions, where the intensity of light becomes very high. At the
focusing regions, neighboring null geodesics meet, and the cross sectional
area of the bundle of light-rays collapses to zero, which leads to the
increase in intensity. Spherical wavefronts focus at a single point, (which
are unstable under small perturbations of the front) whereas generic
wavefronts trace spatial curves of focusing points (cusp ridges and
swallowtails).

In Section II we will review a beautiful method for giving the general
solution to$\ $the flat-space eikonal; it allows for solutions either from
arbitrary Cauchy data or for stationary solutions that arise from the
ansatz, $S=t-S_{0}(x,y,z).$

In Sec. III, we will apply the method of Sec. II to obtain global
asymptotically spherical null surfaces that are associated with shearing
(``bad'') cuts of null infinity\cite{PR},\cite{NT}. They will be defined
from the normal rays to a ``bad'' cut. This construction can be thought of
as beginning with a deformed, initial, two-sphere in a finite region of
space-time, then construct the future outward directed null normals to the
two surface which generates a null surface and finally ``slide'' the initial
two-surface along the null geodesics that generate the null surface, to
future null infinity. This limit is the ``bad'' cut of null infinity.

In Sec. IV, we will study the caustics and singularities of these
characteristic surfaces and their associated wavefronts.

In Sec. V, we treat the same problems of the singularities of these surfaces
but now from an alternate point of view, namely from Arnold's theory of
generating families\cite{A4}. This treatment allows one to handle
(parametrically) the regions of self-intersection and non-smoothness of the
null surfaces.

In Sec. VI, we discuss a generalization of the ideas presented to this
point. Though this generalization is primarily intended for use in non-flat
Lorentzian space-times, nevertheless we believe that it is quite useful to
see it in the simpler case of flat space-time; it allows for the
clarification of certain points that would be difficult in more general
situations. Specifically we will consider solutions of the eikonal equation
that depend on two parameters - that are different from the two parameter
family of plane waves. We will see, the slightly surprising result, that the
singularity structure of the individual characteristic surfaces can be
studied via the parameter behavior of nearby solutions. More precisely, if
the two parameter set of solutions is given by $Z(x^{a},\mu ,\overline{\mu }%
),$ the singularities of the level surfaces of $Z$ for fixed values of ($\mu
,\overline{\mu })$ can be studied and expressed in terms of ($\mu ,\overline{%
\mu })$ derivatives of $Z.$ These results become important in asymptotically
flat space-times where the $Z(x^{a},\mu ,\overline{\mu })$ can be chosen to
represent the family of past null cones from all the points of future null
infinity.

\section{Solutions of the Eikonal Equation}

We review a powerful method for solving the flat-space eikonal equation with
arbitrary given Cauchy data. We begin with a solution $S^{*}$ of the eikonal
equation that depends on three arbitrary parameters, i.e., 
\begin{equation}
S^{*}=S^{*}(x^{i},t,\alpha _{i})=x^{i}\alpha _{i}-t\sqrt{\Sigma (\alpha
_{i})^{2}}  \label{d}
\end{equation}

\noindent called a complete integral. A ``general integral'' (which involves
an arbitrary function) can be constructed from the complete integral in the
following manner: we first add to it an arbitrary function of the three $%
\alpha _{i}$ , i.e., we consider

\begin{equation}
S^{**}=S^{*}(x^{i},t,\alpha _{i})-F(\alpha _{i}),  \label{12}
\end{equation}
with the weak condition that (aside from lower dimensional regions)

\begin{equation}
\left| \frac{\partial ^{2}S^{**}}{\partial \alpha _{i}\partial \alpha _{j}}
\right| \neq 0.
\end{equation}

\noindent We next demand that $\partial S^{**}/\partial \alpha _{i}=\partial
S^{*}/\partial \alpha _{i}-\partial F/\partial \alpha _{i}=0,$ which implies
that there are three functions of the form $\alpha _{i}=A_{i}(x^{i},t)$. (In
general these solutions are not unique and they must be expressed on
different sheets. See Sec. IV for a complete discussion of this issue.)
Finally, via $\alpha _{i}=A_{i}(x^{i},t)$, the $\alpha _{i}$ are eliminated
in the $S^{**}$ yielding (perhaps multivalued)

\begin{equation}
S(x^{i},t)=S^{*}(x^{i},t,A_{i}(x^{i},t))-F(A_{i}(x^{i},t)).  \label{00}
\end{equation}

The level surfaces of this $S$ might self-intersect and be only piece-wise
differentiable.

It is not difficult to show that the $S,$ so constructed satisfies the
eikonal equation \cite{LL}. This follows immediately from the fact that

$\qquad \qquad \partial _{a}S=\partial _{a}S^{*}$ +($\partial _{A_{i}}S^{*}-$
$\partial _{A_{i}}F)\partial _{a}A_{i}=$ $\partial _{a}S^{*}.$

This solution now depends on an arbitrary function of three variables,
namely the $F$. The task is now to determine the $F(\alpha _{i})$ in terms
of (appropriate) Cauchy data, $S_{C}$( $x^{i}$). This is accomplished as
follows; \textsl{\ define} $\alpha _{i}$ = $\partial $ $S_{C}$/$\partial $ $%
x^{i}$ and algebraically invert it in the form of the three equations $x^{i}$
= $X^{i}(\alpha _{i}).$ At $t=t_{0}$ we have that 
\begin{equation}
S(x^{i},t_{0})=S^{*}(x^{i},t_{0},A_{i}(x^{i},t_{0}))-F(A_{i}(x^{i},t_{0})).
\label{13}
\end{equation}
Replacing all the $A_{i}\ $by $\alpha _{i}$ and all the $x^{i}$ by $%
X^{i}(\alpha _{i})$, we have that

\begin{equation}
F(\alpha _{i})=S^{*}(X^{i},t_{0},\alpha _{i})-S_{C}(X^{i}),  \label{14}
\end{equation}
i.e., the free $F(\alpha _{i})$ is now expressed in terms of the free Cauchy
data, $S_{C}(x^{i})$.

This allows us to find (in principle - modulo algebraic inversions or
parametrizations) solutions of the flat-space eikonal equation with
arbitrary Cauchy data.

There exists a special class of solutions that are not studied or found
easily via the Cauchy problem, namely the ``stationary'' solutions which
have the form\qquad $\qquad \qquad \qquad $%
\[
S=t-S_{0}(x^{i}). 
\]
To generate solutions of this form we modify the complete integral, Eq.\ref
{d}, making it a function of only two free parameters by imposing the
condition that $\Sigma (\alpha _{i})^{2}=constant;$ (for convenience chosen
as $\Sigma (\alpha _{i})^{2}=$ $1/\sqrt{2}).$\qquad

We then write the modified complete integral as$\qquad $%
\begin{equation}
S^{*}=x^{a}\ell _{a}(\zeta ,\overline{\zeta })  \label{e}
\end{equation}
where

$\qquad \qquad $%
\begin{equation}
\ell _{a}(\zeta ,\overline{\zeta })=\frac{1}{\sqrt{2}(1+\zeta \overline{%
\zeta })}\left[ (1+\zeta \overline{\zeta }),-(\zeta +\overline{\zeta }),-i(%
\overline{\zeta }-\zeta ),(1-\zeta \overline{\zeta })\right]  \label{ee}
\end{equation}

The complex number $\zeta $, which plays the role of two of the independent
parameters among the three $\alpha _{i},$ can be thought of as the complex
stereographic coordinate on the sphere; the $\ell _{a}(\zeta ,\overline{
\zeta })$ is a Minkowski null vector, $\eta ^{ab}\ell _{a}\ell _{b}=0,$ that
spans the entire lightcone as $\zeta $ ranges over the sphere. Eq.\ref{e}
represents a spheres worth of different families of plane waves parametrized
by the direction $\zeta .$

\qquad If we now take 
\begin{equation}
S^{**}=x^{a}\ell _{a}(\zeta ,\overline{\zeta })+\alpha (\zeta ,\overline{
\zeta })  \label{g}
\end{equation}
and construct \dh $S^{**}=\overline{\text{\dh }}S^{**}$ $=0,$ i.e.,

\begin{eqnarray}
\omega &\equiv &x^{a}m_{a}(\zeta ,\overline{\zeta })+\text{\dh }\alpha
(\zeta ,\overline{\zeta })=0,  \label{h} \\
\text{ }\overline{\omega } &\equiv &x^{a}\overline{m}_{a}(\zeta ,\overline{
\zeta })+\overline{\text{\dh }}\alpha (\zeta ,\overline{\zeta })=0
\label{hh}
\end{eqnarray}
where

\begin{eqnarray}
m_{a}(\zeta ,\overline{\zeta }) &=&\text{\dh }\ell _{a}(\zeta ,\overline{
\zeta })\equiv (1+\zeta \overline{\zeta })\frac{\partial \ell _{a}(\zeta ,%
\overline{\zeta })}{\partial \zeta } \\
\overline{m}_{a}(\zeta ,\overline{\zeta }) &=&\overline{\text{\dh }}\ell
_{a}(\zeta ,\overline{\zeta })\equiv (1+\zeta \overline{\zeta })\frac{%
\partial \ell _{a}(\zeta ,\overline{\zeta })}{\partial \overline{\zeta }}.
\end{eqnarray}

For generic $\alpha (\zeta ,\overline{\zeta }),$ Eq.(\ref{h}) can be solved
for

$\qquad \qquad \qquad \qquad \qquad $%
\begin{equation}
\zeta =\Upsilon (x,y,z),
\end{equation}

\noindent where again these solutions need not be unique and must be
expressed on different sheets. (See Sec. IV for a full treatment of this
problem.) Note that Eqs.(\ref{h}) do not depend on the time $t$ and hence $%
\Upsilon $ is a function only of the spatial coordinates. When the $\Upsilon
(x,y,z)$ is substituted into Eq.(\ref{g}), i.e.,

\begin{equation}
S(t,x,y,z)=\frac{t}{\sqrt{2}}-S_{0}(x,y,z)=\frac{t}{\sqrt{2}}+x^{i}\ell
_{i}(\Upsilon ,\overline{\Upsilon })+\alpha (\Upsilon ,\overline{\Upsilon }),
\label{i}
\end{equation}

\noindent we have a solution of the eikonal equation depending on an
arbitrary function of two variables, $\alpha (\Upsilon ,\overline{\Upsilon }
).$ The level surfaces of $S$ could in general self-intersect and be only
piecewise differentiable.\qquad

The procedure of beginning with a complete solution and obtaining the
general solution via the two (or three) variable arbitrary function $is$
geometrically equivalent to the construction of an envelope from the family
of plane waves as the two (or three) constants in the complete solutions are
varied\cite{CH}.

\qquad Since in this work, we will only be interested in individual null
surfaces and their properties, we can and will confine ourselves to the
level surfaces of the solutions of the form given in Eq.(\ref{i}).

\section{Null Surfaces Generated by Normals to Two-Surfaces \label{2}}

We want to give a slightly different geometric interpretation to the method
of the previous section for generating the stationary solutions of the
eikonal equation. Given any (spatial) two-surface (for example consider any
two dimensional slice of the past light cone of an arbitrary space-time
point), the normal rays to the surface (either the outgoing or incoming
ones) generate a null surface. In this section we will consider a particular
case of this construction where this \textit{past lightcone }is taken to be
the future null infinity, $\frak{I}^{+}$, of Penrose \cite{PR}.

The future null boundary, $\frak{I}^{+}$, (the endpoints of future directed
null geodesics) of any asymptotically simple space-time is a null surface
with topology $R\times S^{2}.$ A choice of Bondi coordinates $(u,\zeta ,\bar{
\zeta})$ can be assigned to $\frak{I}^{+},$ where $\zeta =\cot (\theta
/2)e^{i\phi }$ for the $S^{2}$ sector. The intersection of the future
lightcone $\frak{C}_{x}$ of a point $x^{a}$ with $\frak{I}^{+}$ is a
two-surface, locally imbedded in $R\times S^{2}$; it can generically be
described locally by $u=Z(x^{a},\zeta ,\bar{\zeta})$. The two-surface is
referred to as a lightcone cut\cite{KzN}, whereas the function $%
Z(x^{a},\zeta ,\bar{\zeta})$ is referred to as a lightcone cut function, and
is a two-point real function on the space-time and the boundary, $\frak{I}
^{+}$.

\begin{remark}
Though for this work it is irrelevant, we mention that the lightcone cut
function $Z(x^{a},\zeta ,\bar{\zeta})$ is one of two fundamental variables
in a reformulation of general relativity via null ~surfaces\cite{NSF1,NSF2}.
It encodes all the conformal information of the space-time..
\end{remark}

In the remainder of this work we will confine ourselves to flat space-time
where (modulo Poincare transformations) a natural choice of Bondi
coordinates $(u_{n},\zeta ,\bar{\zeta})$ exists; the $u_{n}$ $=constant$ is
constructed from the intersection of the future lightcone, $\frak{C}%
_{(t,o,o,o)}$, of the spatial origin, at time $t=u_{n},$ with $\frak{I}^{+};$
the ($\zeta ,\bar{\zeta})$ are just the null directions, at the origin,
carried along by the null generators of the lightcone. Using Cartesian
coordinates $x^{a}$ for the space-time and these natural Bondi coordinates,%
\cite{KN} the lightcone cuts can be described as 
\begin{equation}
u_{n}=x^{a}\ell _{a}(\zeta ,\bar{\zeta})  \label{natural}
\end{equation}
where $\ell _{a}$ represents the covariant version of a null vector $\ell
^{a}$ with Cartesian components given as 
\begin{eqnarray}
\ell ^{a}(\zeta ,\bar{\zeta}) &=&\frac{1}{\sqrt{2}(1+\zeta \bar{\zeta})}%
\left( (1+\zeta \bar{\zeta}),(\zeta +\bar{\zeta}),i(\bar{\zeta}-\zeta
),(\zeta \bar{\zeta}-1)\right)  \label{nn} \\
&=&\frac{1}{\sqrt{2}}\left( 1,\sin \theta \cos \phi ,\sin \theta \sin \phi
,\cos \theta \right) .  \nonumber
\end{eqnarray}

Adding the radial coordinate $r,$ this natural choice of Bondi coordinates
is identical to the standard null polar coordinates $(u_{n},r,\zeta ,\bar{
\zeta})$ given by\qquad 
\begin{equation}
x^{a}=ut^{a}+r\ell ^{a}(\zeta ,\bar{\zeta}),\text{ }t^{a}\text{ = }\sqrt{2}%
\text{(1,0,0,0)}
\end{equation}

Note that Eqs.(\ref{natural}) and (\ref{nn}) are identical with Eqs.(\ref{e}
) and (\ref{ee}) though their meanings are different; Eq.(\ref{natural}) has
the dual meaning of being, for fixed value of the $x^{a},$ the lightcone cut
of $\frak{I}^{+}$ and also, for fixed values of $(u_{n},\zeta ,\bar{\zeta}),$
it describes the plane wave (null surface) intersecting the time axis at $%
t=u_{n}$ in the direction of $(\zeta ,\bar{\zeta}).$

By Eq.(\ref{natural}), the lightcone cuts of any points $x^{a}=(t,0,0,0)$
along the time axis take, as we mentioned earlier, a constant value on $%
\frak{I}^{+}$, namely, they are the constant-$u_{n}$ slices. The natural
Bondi cuts are lightcone cuts as well. By following inwardly the null
geodesics that leave the natural Bondi cuts orthogonally, we find no
focusing other than at the apex (on the time axis) of the lightcone.

By a slight modification of the above we can find other null surfaces
leaving $\frak{I}^{+}$ that have much more complicated focusing properties
than that of a lightcone. If we consider the one parameter family of cuts of 
$\frak{I}^{+}$ given, say by, $\qquad \qquad \qquad \qquad $%
\begin{equation}
u_{n}=-\alpha (\zeta ,\bar{\zeta})+u  \label{j}
\end{equation}
\noindent where $\alpha (\zeta ,\bar{\zeta})$ is a given but arbitrary
regular function on $S^{2},$ and $u$ is a parameter on $R,$ we can ask for
the null surfaces generated by the null normals to the family of cuts.

\begin{itemize}
\item  Note that Eq.(\ref{j}) can be rewritten as $u=u_{n}+\alpha (\zeta ,%
\bar{\zeta})$ and reinterpreted as a (Bondi-Metzner-Sachs) supertranslation%
\cite{NT} between the coordinates $u_{n}$ and $u$ on $\frak{I}^{+}.$
\noindent \noindent 
\end{itemize}

We now construct the null surface formed by the normal rays to the cuts, Eq.(%
\ref{j}), determined by $u=const.;$ replacing the $u_{n}$ in Eq$.(\ref{j})$
by the null planes Eq.(\ref{natural}), [$u_{n}=x^{a}\ell _{a}(\zeta ,\bar{
\zeta})],$ we have$\qquad \qquad \qquad $%
\begin{equation}
u=x^{a}\ell _{a}(\zeta ,\bar{\zeta})+\alpha (\zeta ,\bar{\zeta})  \label{k}
\end{equation}
which is identical to Eq.(\ref{g}). The envelope formed from all the null
planes that are normal to the family of cuts are found by setting to zero
the $\zeta $ and $\bar{\zeta}$ derivatives of Eq.(\ref{k}) and eliminating
the ($\zeta ,\bar{\zeta})$ from (\ref{k}), a procedure \textit{identical }
to that followed in the previous section to obtain Eq.(\ref{i}), i.e., we
now have the one parameter family of null surfaces

\qquad 
\begin{equation}
S^{**}\equiv u=\frac{t}{\sqrt{2}}-x^{i}\ell _{i}(\Upsilon ,\overline{
\Upsilon })+\alpha (\Upsilon ,\overline{\Upsilon })  \label{l}
\end{equation}
with $\zeta =\Upsilon (x,y,z),$ a solution of Eqs.(\ref{h}),(\ref{hh}). The
procedure of setting to zero the $\omega $ and $\overline{\omega }$ in

\begin{eqnarray}
\omega  &\equiv &x^{a}m_{a}(\zeta ,\overline{\zeta })+\text{\dh }\alpha
(\zeta ,\overline{\zeta })=0,  \label{ll} \\
\text{ }\overline{\omega } &\equiv &x^{a}\overline{m}_{a}(\zeta ,\overline{%
\zeta })+\overline{\text{\dh }}\alpha (\zeta ,\overline{\zeta })=0
\label{lll}
\end{eqnarray}
selects the null ray at each point of $\frak{I}^{+}$ that is orthogonal to
the cut given by Eq.(\ref{j}).

\begin{itemize}
\item  Note that $\alpha $ can be chosen to contain only spherical harmonics
of order $l\eqslantgtr 2$ since any $l=0,1$ components of $\alpha $ could be
absorbed by $x^{a}\ell _{a}$ with no modification other than displacing the
origin of the coordinates $x^{a}$, since $\ell _{a}$ is precisely the
collection of spherical harmonics of order 0 and 1.
\end{itemize}

\noindent \qquad We can give a parametric description of the family of null
surfaces, Eq.(\ref{l}), by the following procedure: we consider the four
functions

\begin{equation}
\zeta =\Upsilon (x,y,z)\Leftrightarrow x^{a}m_{a}(\zeta ,\overline{\zeta })+%
\text{\dh }\alpha (\zeta ,\overline{\zeta })=0  \label{m}
\end{equation}

\begin{equation}
\overline{\zeta }=\overline{\Upsilon }(x,y,z)\Leftrightarrow x^{a}\overline{m%
}_{a}(\zeta ,\overline{\zeta })+\overline{\text{\dh }}\alpha (\zeta ,%
\overline{\zeta })=0  \label{mm}
\end{equation}

\begin{equation}
u=u(t,x,y,z)=\{x^{a}\ell _{a}(\zeta ,\bar{\zeta})+\alpha (\zeta ,\bar{\zeta}
)\}|_{\Upsilon \overline{,\Upsilon }}  \label{mmm}
\end{equation}

\begin{equation}
r=r(x,y,z)=\{x^{a}(n_{a}-\ell _{a})(\zeta ,\bar{\zeta})+\text{\dh }\overline{%
\text{\dh }}\alpha (\zeta ,\bar{\zeta})\}|_{\Upsilon \overline{,\Upsilon }}
\label{4m}
\end{equation}
and consider them as a coordinate transformation between the \{$u,\zeta ,%
\overline{\zeta },r\}$ and the \{$x^{a}\}.$ We have used $m^{a}\equiv $ \dh $%
\ell ^{a}$, $\bar{m}^{a}\equiv $ $\overline{\text{\dh }}\ell ^{a}$ , and $%
n^{a}\equiv $ \dh $\overline{\text{\dh }}\ell ^{a}+\ell ^{a}$. From the fact
that $(\ell ^{a},m^{a},\bar{m}^{a},n^{a})$ form a null tetrad for every
fixed value of $(\zeta ,\bar{\zeta})$, this coordinate transformation can be
readily inverted into the form 
\begin{equation}
x^{a}=(u-\alpha )(n^{a}+\ell ^{a})+(r-\text{\dh }\overline{\text{\dh }}
\alpha )\ell ^{a}+(\text{\dh }\alpha )\bar{m}^{a}+(\overline{\text{\dh }}
\alpha )m^{a}  \label{n}
\end{equation}
This relationship can alternatively be looked on as the parametric version
of the one-parameter family of null surfaces, Eq.(\ref{l}), (where, for
fixed $u,$ the $(r$,$\zeta ,\bar{\zeta})$ parametrize the surface) or as the
coordinate transformation between the \{$x^{a}\}$ and the null-geodesic
coordinates, $(u,\zeta ,\bar{\zeta},r)$; $u$ labels the null surfaces, the
pair $(\zeta ,\bar{\zeta})$ labels null geodesics (via their intersection
with $\frak{I}^{+})$ and $r$ is an affine parameter along the null
geodesics. That this is true can be easily seen from the parametric form,
Eq.(\ref{n}), by simply constructing

\qquad \qquad 
\begin{equation}
\frac{dx^{a}}{dr}=\ell ^{a}(\zeta ,\bar{\zeta})  \label{o}
\end{equation}
and observing that $\ell ^{a}$ is a null tangent vector with affine
normalization.

The transformation between the $x^{a}$ and the $(u,r,\zeta ,\bar{\zeta})$
breaks down when the Jacobian of the transformation, Eq.(\ref{n}), vanishes,
i.e., when 
\begin{equation}
D\equiv \frac{\partial (t,x,y,z)}{\partial (u,r,\zeta ,\bar{\zeta})}=r^{2}-%
\text{\dh }^{2}\alpha \text{ }\overline{\text{\dh }}^{2}\alpha =0.
\label{oo}
\end{equation}

\noindent Geometrically, this is where the null surface develops
singularities. In the projection to the three-space $(x,y,z)$ it is a two
surface; the ``caustic surface''. To see this explicitly, we return to Eq.(%
\ref{n}) where we have (for fixed $u$) that $(x,y,z)=x^{i}=$ $X^{i}(r,\zeta ,%
\bar{\zeta}),$ i.e., are known functions of $(r,\zeta ,\bar{\zeta}).$ If the 
$r$ in $X^{i}$ is replaced by the $r$ from Eq.(\ref{oo}) we have the
parametric form of the caustic, $\qquad \qquad \qquad $%
\begin{equation}
x^{i}=X^{i}(r(\zeta ,\bar{\zeta}),\zeta ,\bar{\zeta})=\widehat{X}^{i}(\zeta ,%
\bar{\zeta}).  \label{ooo}
\end{equation}
We will return to this in the next section.

It is interesting to note that the coordinates $(u,\zeta ,\bar{\zeta},r)$
represent a type of null coordinate system that we could call \textit{\
asymptotic null-polar coordinates} which are the flat space case of an \emph{%
\ \ \ \ \ interior Bondi coordinate system\/}~\cite{FKN,NT}, i.e., the
extension into the interior of the space-time of the Bondi coordinates $(u$,$%
\zeta ,\bar{\zeta})$ on $\frak{I}^{+}.$ They differ from the standard null
polar coordinates by the fact that the null geodesics that rule these
surfaces possess non-vanishing shear while for the standard ones the shear
vanishes.

The complex shear is defined as $\sigma =$ $M^{a}M^{b}\nabla _{a}L_{b},$
where $L_{b}$ is tangent to the null geodesics and $M^{a}$ is complex null,
orthogonal to $L_{b}$ and such that $M^{a}\overline{M}_{a}=-1.$ In our case,
because of Eqs.(\ref{n}) and (\ref{o}) and the fact that $(\ell ^{a},m^{a},%
\bar{m}^{a},n^{a})$ forms a null tetrad, we have that $L_{b}=\ell _{b}$ and $%
M^{a}=m^{a}.$ Furthermore, the gradient of $\ell _{b}\ $is $\nabla _{a}\ell
_{b}=$ $m_{a}\zeta ,_{b}+\overline{m}_{a}\overline{\zeta },_{b}$ and thus $%
\sigma =-m^{b}\overline{\zeta },_{b}.$ To obtain the derivative of $%
\overline{\zeta }$ along $m^{b}$ we take the gradients of Eqs.(\ref{ll}) and
(\ref{lll}) which yields

\begin{eqnarray}
m_{b}(\zeta ,\overline{\zeta })+\{x^{a}(n_{a}-l_{a})+\text{\dh }\overline{%
\text{\dh }}\alpha \}\overline{\zeta },_{b}+\text{\dh }^{2}\alpha \zeta
,_{b} &=&0,  \label{2o} \\
\overline{m}_{b}(\zeta ,\overline{\zeta })+\{x^{a}(n_{a}-l_{a})+\text{\dh }%
\overline{\text{\dh }}\alpha \}\zeta ,_{b}+\overline{\text{\dh }}^{2}\alpha 
\overline{\zeta },_{b} &=&0.  \label{3o}
\end{eqnarray}
Using Eq.(\ref{4m}) and contracting Eqs.(\ref{2o}) and (\ref{3o}) by $m^{b}$
, we obtain

\begin{eqnarray}
r\,m^{b}\overline{\zeta },_{b}+\text{\dh }^{2}\alpha \,m^{b}\zeta ,_{b} &=&0,
\\
-1+r\,m^{b}\zeta ,_{b}+\overline{\text{\dh }}^{2}\alpha \,m^{b}\overline{%
\zeta },_{b} &=&0.
\end{eqnarray}
By eliminating the $m^{b}\zeta ,_{b}$ from these equations, we find

\begin{equation}
(r^{2}-\text{\dh }^{2}\alpha \overline{\,\text{\dh }}^{2}\alpha )m^{b}%
\overline{\zeta },_{b}+\text{\dh }^{2}\alpha =0
\end{equation}
or\qquad 
\begin{equation}
\sigma =\frac{\sigma ^{0}}{r^{2}-\sigma ^{0}\overline{\sigma }^{0}}.
\label{p}
\end{equation}
where $\sigma ^{0}=$ \dh $^{2}\alpha .$ This is also a confirmation of the
Sachs theorem on the transformation of the asymptotic shear, $\sigma ^{0},$
under a BMS transformation. Eq.(\ref{p}) represents a special (non-twisting)
case of a more general result valid for generic null congruences in flat
space\cite{LN}.

We point out that the flat-space line element, using Eq.(\ref{n}), can
easily be expressed in terms of these shearing Bondi coordinates as\cite
{Wetal}

\begin{eqnarray}
ds^{2} &=&\eta _{ab}dx^{a}dx^{b}  \nonumber \\
&=&2du\left( du+dr-\bar{\text{\dh }}\text{\dh }^{2}\alpha \frac{d\zeta }{P}-%
\text{\dh }\overline{\text{\dh }}^{2}\alpha \frac{d\bar{\zeta}}{P}\right) 
\nonumber \\
&&{}-\frac{2r}{P^{2}}(\text{\dh }^{2}\!\alpha \,d\zeta ^{2}+\overline{\text{%
\dh }}^{2}\!\alpha \,d\bar{\zeta}^{2})-2(r^{2}+\text{\dh }^{2}\!\alpha \,%
\overline{\text{\dh }}^{2}\!\alpha )\frac{d\zeta d\bar{\zeta}}{P^{2}},
\label{q}
\end{eqnarray}
where $P=1+\zeta \bar{\zeta}$ . The determinant of $g_{ab}$ is given by $%
\left| g\right| =\frac{1}{P^{4}}\{r^{2}-$ \dh $^{2}\alpha \overline{\,\text{
\dh }}^{2}\alpha \}^{2}=$ $\frac{D^{2}}{P^{4}}$ whose vanishing agrees with
the vanishing of the Jacobian, Eq.(\ref{oo}).

Note that the asymptotic $r=const.\Rightarrow \infty $ sections, at $%
u=const.,$ become metric spheres.

\begin{itemize}
\item  It is perhaps interesting to speculate on the use of Eq. (\ref{q}) as
the Minkowski space lowest order term, in perturbation calculations, for
solutions of the Einstein equations.
\end{itemize}

We complete this section by showing how a null surface can be constructed
explicitly from the normals to an arbitrary space-like two surface, $\frak{S}
$, in a manner virtually identical to those constructed from a cut or slice
of null infinity.

We begin from Eq.(\ref{k})

$\qquad \qquad \qquad u=0=x^{a}\ell _{a}(\zeta ,\bar{\zeta})+\alpha (\zeta ,%
\bar{\zeta})$

\noindent with Eqs. (\ref{ll} and \ref{lll}). The issue is, given the
surface $\frak{S}$, how is one to choose $\alpha (\zeta ,\bar{\zeta})$?

First $\frak{S}$ is defined parametrically, $x^{a}=x_{0}^{a}(\zeta ,\bar{%
\zeta}),$where the parameters are chosen as follows: consider a time-like
world line at the spacial origin and the family of light-cones centered on
the line. The null geodesics ruling these cones are labeled by their
directions ($\zeta ,\bar{\zeta})$ on the sphere and coincide with labeling
of the generators of null infinity. The points on $\frak{S}$ are now
(locally) parametrized by the labels of the null geodesics passing thru
those points. With this parametrization the function $\alpha (\zeta ,\bar{%
\zeta})$ is defined by

$\qquad \qquad \alpha (\zeta ,\bar{\zeta})=-x_{0}^{a}(\zeta ,\bar{\zeta}
)\ell _{a}(\zeta ,\bar{\zeta})$

\noindent so that Eq.(\ref{l}) becomes

$\qquad u=0=(x^{a}-x_{0}^{a}(\zeta ,\bar{\zeta}))\ell _{a}(\zeta ,\bar{\zeta}
)$

\noindent and Eqs.(\ref{ll} and \ref{lll}) become

\begin{equation}
(x^{a}-x_{0}^{a}(\zeta ,\bar{\zeta}))m_{a}(\zeta ,\bar{\zeta})-\ell
_{a}(\zeta ,\bar{\zeta})\text{\dh }x_{0}^{a}=0\text{ and }c.c.  \label{pp}
\end{equation}

We see that the null surface so defined goes thru $\frak{S,}$ i.e.$,$ thru $%
x^{a}=x_{0}^{a}(\zeta ,\bar{\zeta})$. To see that it is also normal to $%
\frak{S}$, we notice that at $\frak{S}$ the first two terms of Eq.(\ref{pp})
cancel out and we are left at $\frak{S}$ with

$\qquad \qquad \qquad \ell _{a}(\zeta ,\bar{\zeta})$ \dh $x_{0}^{a}=0.$

Thus as was claimed, the tangent vectors to $\frak{S}$, namely \dh $%
x_{0}^{a}(\zeta ,\bar{\zeta})$, are normal to the null tangent vectors to
the null surface, $\ell _{a}.$

We see that the earlier construction of null surfaces from cuts of null
infinity actually includes those constructed from finite surfaces.

\section{Wavefront evolution and singularities \label{3}}

In the previous section, we mentioned that the null coordinate system broke
down and the associated shearing null surfaces developed caustics at the
points where $r^{2}=$ \dh $^{2}\alpha $ $\overline{\text{\dh }}^{2}\alpha .$

Here we focus our attention on the two dimensional wavefronts associated
with these null surfaces. We show that the wavefronts develop singularities,
and we locate the singularities via the standard method of singularity
theory, and via our lightcone cut approach. The evolution of these
singularities as the wave fronts evolve become the caustics.

A wavefront is, by definition, the intersection of our null surface, $%
u=u_{0},$ with a constant-time $t_{0}$ surface. This represents an instant
in the progression of a wave. In our case, this requires fixing the time
coordinate $x^{0}=t_{0}$ in Eq.(\ref{n}),~and solving for

\begin{equation}
r=\sqrt{2}t_{0}-2u_{0}+2\alpha +\text{\dh }\overline{\text{\dh }}\alpha .
\label{qq}
\end{equation}
The remaining coordinates $(x,y,z)$, using Eq.(\ref{qq}) to eliminate $r$,
trace a two-surface (a ``small'' wave front) in the Euclidean three-space,
parametrized by $(\zeta ,\bar{\zeta})$ [or $(\theta ,\phi )$ under the
transformation $\zeta =e^{i\phi }\cot {\theta /2}$]:

\begin{eqnarray}
x &=&\frac{1}{\sqrt{2}}\left[ (\sqrt{2}t_{0}-2u_{0}+2\alpha )\frac{(\zeta +%
\bar{\zeta})}{(1+\zeta \bar{\zeta})}+\bar{\text{\dh }}\alpha \frac{(1-\bar{%
\zeta}^{2})}{(1+\zeta \bar{\zeta})}+\text{\dh }\alpha \frac{(1-\zeta ^{2})}{%
(1+\zeta \bar{\zeta})}\right] \\
y &=&\frac{i}{\sqrt{2}}\left[ (\sqrt{2}t_{0}-2u_{0}+2\alpha )\frac{(%
\overline{\zeta }-\zeta )}{(1+\zeta \bar{\zeta})}-\bar{\text{\dh }}\alpha 
\frac{(1+\bar{\zeta}^{2})}{(1+\zeta \bar{\zeta})}+\text{\dh }\alpha \frac{%
(1+\zeta ^{2})}{(1+\zeta \bar{\zeta})}\right] \\
z &=&\frac{1}{\sqrt{2}}\left[ (\sqrt{2}t_{0}-2u_{0}+2\alpha )\frac{(\zeta 
\bar{\zeta}-1)}{(1+\zeta \bar{\zeta})}+\text{\={\dh }}\alpha \frac{2\bar{%
\zeta}}{(1+\zeta \bar{\zeta})}+\text{\dh }\alpha \frac{2\zeta }{(1+\zeta 
\bar{\zeta})}\right]
\end{eqnarray}
The map $(\zeta ,\bar{\zeta})\rightarrow (x,y,z)$ is singular at points
where the Jacobian matrix 
\begin{equation}
\left( 
\begin{array}{ll}
\text{\dh }x & \bar{\text{\dh }}x \\ 
\text{\dh }y & \bar{\text{\dh }}y \\ 
\text{\dh }z & \bar{\text{\dh }}z
\end{array}
\right)
\end{equation}
drops rank, from 2 to 1 or 0. The drop in rank takes place if the three
2-determinants vanish simultaneously: 
\begin{eqnarray}
\text{\dh }x\bar{\text{\dh }}y-\text{\dh }y\bar{\text{\dh }}x &=&0, \\
\text{\dh }y\bar{\text{\dh }}z-\text{\dh }z\bar{\text{\dh }}y &=&0, \\
\text{\dh }z\bar{\text{\dh }}x-\text{\dh }x\bar{\text{\dh }}z &=&0.
\end{eqnarray}
Since 
\begin{eqnarray}
\text{\dh }x^{a}|_{u,t} &=&\text{\dh }^{2}\alpha \bar{m}^{a}+rm^{a}, \\
\bar{\text{\dh }}x^{a}|_{u,t} &=&\bar{\text{\dh }}^{2}\alpha m^{a}+r\bar{m}%
^{a},
\end{eqnarray}
the explicit expressions of the 2-determinants are as follows 
\begin{eqnarray}
\text{\dh }x\bar{\text{\dh }}y-\text{\dh }y\bar{\text{\dh }}x &=&(r^{2}-%
\text{\dh }^{2}\alpha \bar{\text{\dh }}^{2}\alpha )(m^{x}\bar{m}^{y}-\bar{m}%
^{x}m^{y}), \\
\text{\dh }y\bar{\text{\dh }}z-\text{\dh }z\bar{\text{\dh }}y &=&(r^{2}-%
\text{\dh }^{2}\alpha \bar{\text{\dh }}^{2}\alpha )(m^{y}\bar{m}^{z}-\bar{m}%
^{y}m^{z}), \\
\text{\dh }z\bar{\text{\dh }}x-\text{\dh }x\bar{\text{\dh }}z &=&(r^{2}-%
\text{\dh }^{2}\alpha \bar{\text{\dh }}^{2}\alpha )(m^{z}\bar{m}^{x}-\bar{m}%
^{z}m^{x}).
\end{eqnarray}
Thus, all three determinants vanish at points where 
\begin{equation}
D\equiv r^{2}-\text{\dh }^{2}\alpha \bar{\text{\dh }}^{2}\alpha =0
\label{area}
\end{equation}
which, with $r$ from Eq.(\ref{qq}), determines a curve, the wavefront
singularities. Note that the evolution of the wavefront singularities
(obtained by varying $t_{0})$ yields the caustic surface.

Since $\alpha $ is a regular function on the sphere, so is \dh $^{2}\alpha $ 
$\overline{\text{\dh }}^{2}\alpha $; therefore, Eq.(\ref{area}) (with $r$
given by Eq.(\ref{qq})) admits solutions $\zeta $ only within a finite
interval of time $t$. Thus the wavefronts are singular only during a closed
interval of time. On the other hand, at very long times the wavefronts
become spherical, which follows from the line element Eq.(\ref{q}).

The wavefront singularities (curves) are places where neighboring null
geodesics meet. We have a null surface $x^{a}(r,\zeta ,\bar{\zeta})$
foliated by null geodesics. At every fixed value of $r$, there are two
connecting vectors \dh $x^{a}|_{u,r}$ and $\bar{\text{\dh }}x^{a}|_{u,r}$.
The null geodesics in this congruence meet wherever the area orthogonal to
the congruence, spanned by the connecting vectors, vanishes. The connecting
vectors are, explicitly, 
\begin{eqnarray}
\text{\dh }x^{a}|_{u,r} &=&-\text{ }\bar{\text{\dh }}\text{\dh }^{2}\alpha
\ell ^{a}+\text{\dh }^{2}\alpha \bar{m}^{a}+rm^{a}, \\
\bar{\text{\dh }}x^{a}|_{u,r} &=&-\text{ \dh }\overline{\text{\dh }}%
^{2}\alpha \ell ^{a}+r\bar{m}^{a}+\overline{\text{\dh }}^{2}\alpha m^{a},
\end{eqnarray}
The area spanned by the connecting vectors (calculated from their skew
product) is simply $D\equiv r^{2}-$ \dh $^{2}\alpha \overline{\text{\dh }}%
^{2}\alpha $. The vanishing of this area takes place at exactly the points
given by Eq.~(\ref{area}).

We close this section with two examples. \vspace{1cm}

\textbf{Example 1:} $\alpha =Y_{20}=3\cos ^{2}\theta -1$. Due to axial
symmetry, the wavefronts and their singularities for this choice of $\alpha $
can be completely worked out analytically, which gives insights into more
general cases. The wavefronts at a given time $t$ are given by 
\begin{eqnarray}
x &=&\frac{1}{\sqrt{2}}\sin {\theta }\cos {\phi }(\sqrt{2}t-2u-2-6\cos
^{2}\theta ), \\
y &=&\frac{1}{\sqrt{2}}\sin {\theta }\sin {\phi }(\sqrt{2}t-2u-2-6\cos
^{2}\theta ), \\
z &=&\frac{1}{\sqrt{2}}\cos {\theta }(\sqrt{2}t-2u+10-6\cos ^{2}\theta ).
\end{eqnarray}

These are axially symmetric. For a closed interval of time, all the
wavefronts are singular. For early and late times, however, the wavefronts
are smooth.

The singular points are located by eliminating $r$ from Eqs.(\ref{area}) and
(\ref{qq}), yielding in this case, the two solutions or ``sheets''

\begin{equation}
(\sqrt{2}t-2u+10-18\cos ^{2}\theta )(\sqrt{2}t-2u-2-6\cos ^{2}\theta )=0.
\label{areae}
\end{equation}

There is a solution $\theta $ only at times $\sqrt{2}(u-5)\le t\le \sqrt{2}%
(4+u)$. This is the interval where every wavefront is singular. A smooth
wavefront and its corresponding profile at a late time are shown in Fig.~1.
A wavefront at a time when both the cusp ridge singularities and the $z$
-axis singularities are occurring, and its corresponding profile, are shown
in Fig.~2. In Fig. 3, we have a later wavefront and profile with only the
cusp ridge singularity. In Fig. 4, we display the evolution of the
singularities forming the caustic 2-surface.

The high symmetry of this case is responsible for the lack of resemblance of
the singular points on the $z$-axis with standard cusps. At these points,
null geodesics labeled by different, but neighboring values of $\phi ,$
meet. This is clear from the fact that $\partial x/\partial \phi =\partial
x/\partial \phi =\partial z/\partial \phi =0$ at these points, therefore the
vector that connects geodesics with different values of $\phi $ vanishes.

In order to make a comparison, Fig.~5 shows a wavefront in the evolution of
an imploding ellipsoid of revolution which is very similar to that of
Example 1. In this case, an ellipsoid of revolution sends an incoming
wavefront, which develops singularities during a certain interval of time.
The standard cuspoidal ridges are clearly visible as rings at both ends of
the figure. However, the crossover points in between are also singular, of
the same type of singularity as that one developed in our example. Assuming
a speed of light of 1, the formulas for the imploding wavefront in this case
are 
\begin{eqnarray}
x &=&a\sin {\theta }\cos {\phi }\left( 1-\frac{t}{\sqrt{a^{2}\sin ^{2}{%
\theta }+(a^{2}/c)^{2}\cos ^{2}\theta }}\right) , \\
y &=&a\sin {\theta }\sin {\phi }\left( 1-\frac{t}{\sqrt{a^{2}\sin ^{2}{%
\theta }+(a^{2}/c)^{2}\cos ^{2}\theta }}\right) , \\
z &=&c\cos {\theta }\left( 1-\frac{t}{\sqrt{(c^{2}/a)^{2}\sin ^{2}{\theta }%
+c^{2}\cos ^{2}\theta }}\right) .
\end{eqnarray}

\begin{quote}
\textbf{Example 2:} $\alpha =$ Real$(Y_{21})=(\zeta +\bar{\zeta})(\zeta \bar{%
\zeta}-1)/(1+\zeta \bar{\zeta})^{2}$. In this case, there is no advantage in
writing the wavefronts explicitly. However, they can be plotted with ease,
displaying the typical singularities of three-dimensional wavefronts, namely
swallowtails and cusp ridges. Cusp ridges are clearly visible in Fig.~6,
which represents a wavefront at $u=0$ and $t=1.5$. Swallow tails are
exemplified in Fig.7, which represents another wavefront in the same $u=0$
family, but at a later time of $t=2.35$. Locally all swallowtail have the
form of Fig.8.

\qquad Both Figs.6 and 7 compare remarkably well with a typical wavefront in
the evolution of a triaxial ellipsoid, in which case an ellipsoid $%
(x/a)^{2}+(y/b)^{2}+(z/c)^{2}=1$ emits a wavefront of light inwardly, which
develops singularities for a period of time. A typical singular imploding
wavefront is shown in Fig.~9. The formulas for the imploding
triaxial-ellipsoidal wavefront are the following: 
\begin{eqnarray}
x &=&a\sin {\theta }\cos {\phi (}1-\frac{t}{a^{2}\sqrt{(\sin {\theta }\cos {%
\ \ \phi }/a^{2})^{2}+(\sin {\theta }\sin {\phi }/b^{2})^{2}+(\cos {\theta }%
/c^{2})^{2}}})  \nonumber \\
y &=&b\sin {\theta }\sin {\phi (}1-\frac{t}{b^{2}\sqrt{(\sin {\theta }\cos {%
\ \ \ \phi }/a^{2})^{2}+(\sin {\theta }\sin {\phi }/b^{2})^{2}+(\cos {\theta 
}/c^{2})^{2}}})  \nonumber \\
z &=&c\cos {\theta (}1-\frac{t}{c^{2}\sqrt{(\sin {\theta }\cos {\phi }%
/a^{2})^{2}+(\sin {\theta }\sin {\phi }/b^{2})^{2}+(\cos {\theta }/c^{2})^{2}%
}})
\end{eqnarray}
\end{quote}

\section{\noindent Generating Families}

In this section we will study the subject of the caustics of the null
surfaces and the wavefront singularities via an alternative method, namely
from the use of generating families for the construction of Lagrangian and
Legendre submanifolds (developed by V. I. Arnold and his colleagues \cite
{A1,A2,A3,A4}) associated with cotangent and contact bundles over
space-time. The value of this treatment is that it allows one to deal (via a
parametric representation) with the regions of self-intersection and
non-differentiability of the null surfaces.

We first give a brief review of a special case of this theory that is
adapted to the problem of null surfaces in \noindent four-dimensional
space-time. Consider a four dimensional Lorentzian manifold, (with local
coordinates, \{$t,x^{i}\}$) foliated by the constant $t$ surfaces. Now
consider the $x^{i}$ as the coordinates of a configuration space $M$ and $%
p_{i}$ as the conjugate momentum so that we have the six dimensional
cotangent bundle $T^{*}M,$ with local coordinates $(x^{i},p_{i})$. We now
describe the construction of a three dimensional submanifold of $T^{*}M$ (a
Lagrangian submanifold - a maximal submanifold such that the symplectic
form, restricted to it, vanishes) that plays a fundamental role in the
discussion of the singularities of wavefronts and their associated caustics.
We begin with a general description without any particular choice of
dynamics, later restricting ourselves to null geodesic motion.

\qquad Choose a scalar function (determined later from the dynamics),
referred to as a generating family, of the six variables $\qquad \qquad
\qquad \qquad \qquad $%
\begin{equation}
f=F(t,x^{i},\zeta ,\bar{\zeta});  \label{r}
\end{equation}
the $x^{i}$ are spatial points$,$ the $(\zeta ,\bar{\zeta})$ are parametric
labels (for convenience we are using a complex representation) for points on
a given spatial two-surface, $\frak{s,}$ i.e., we have a two-point function,
while $t$ is the time for the (dynamic) particle to go from a point on $%
\frak{s}$ to the point $x^{i}.$ For a constant value of $f,$ we consider Eq.(%
\ref{r}) as defining $t$ implicitly as a function of ($x^{i},\zeta ,\bar{%
\zeta}),$ i.e.,

\qquad \qquad \qquad \qquad \qquad \qquad \qquad \qquad \qquad $%
t=T(f;x^{i},\zeta ,\bar{\zeta})$

\noindent or simply

\begin{equation}
t=T(x^{i},\zeta ,\bar{\zeta}).  \label{rr}
\end{equation}
Note that $T$ might be a multivalued function of its arguments, in which
case it must be considered separately on the different sheets.

We now ask for the relationship between the ($x^{i},\zeta ,\bar{\zeta})$
when $T$ is an extremal under variations of the ($\zeta ,\bar{\zeta});$
i.e., we require that

$\qquad \qquad \qquad \qquad \qquad \qquad \qquad \qquad $%
\begin{equation}
\partial T/\partial \zeta =\partial T/\partial \bar{\zeta}=0  \label{rrr}
\end{equation}
which in turn forces\qquad $\qquad \qquad \qquad \qquad $%
\begin{equation}
\partial F/\partial \zeta =\partial F/\partial \bar{\zeta}=0.  \label{s}
\end{equation}
Finally a rank condition is imposed on the choice of $F;$ the following $2$x$%
5$ matrix must have rank two

\begin{mathletters}
\begin{equation}
\left[ 
\begin{array}{c}
\frac{\partial ^{2}F}{\partial \zeta ^{2}},\quad \frac{\partial ^{2}F}{%
\partial \zeta \partial \overline{\zeta }},\quad \frac{\partial ^{2}F}{%
\partial \zeta \partial x^{i}} \\ 
\frac{\partial ^{2}F}{\partial \zeta \partial \overline{\zeta }},\quad \frac{
\partial ^{2}F}{\partial \overline{\zeta }^{2}},\quad \frac{\partial ^{2}F}{%
\partial \overline{\zeta }\partial x^{i}}
\end{array}
\right] .  \label{t}
\end{equation}
The meaning of this condition is that the two equations (\ref{s} or \ref{rrr}%
) can be solved (locally) in at least one of a variety of possible ways for
two of the five variables ($x^{i},\zeta ,\bar{\zeta})$; often it is
necessary to solve them in the different ways in different regions. We then
have three different possible cases:

\begin{enumerate}
\item  $\zeta =\Upsilon (x^{i}),\bar{\zeta}=\overline{\Upsilon }(x^{i});$
the simplest of the three cases. It allows $F$ to be treated as a function
of just $x^{i}.$ In the other cases $F$ must be treated parametrically.

\item  $\zeta =\Psi (x^{A},\bar{\zeta}),$ $x^{J}=$ $X^{J}(x^{A},\bar{\zeta}),
$ where $x^{A}$ are any two of the three $x^{i}$ and $x^{J}$ is the third
one - or the conjugate version, $\overline{\zeta }=\overline{\Psi }%
(x^{A},\zeta ),$ $x^{J}=$ $X^{J}(x^{A},\zeta ).$

\item  $x^{A}=$ $X^{A}(x^{J},\zeta ,\bar{\zeta}),$ where again $x^{A}$ are
any two of the three $x^{i}$ and $x^{J}$ is the third one.
\end{enumerate}

Case 1 can occur when the determinant

\qquad $\qquad \qquad \qquad $%
\end{mathletters}
\begin{equation}
\hat{D}=\left| 
\begin{array}{ll}
\frac{\text{$\partial $}^{2}F}{\partial \zeta ^{2}} & \frac{\text{$\partial $%
}^{2}F}{\partial \zeta \partial \overline{\zeta }} \\ 
\frac{\text{$\partial $}^{2}F}{\partial \zeta \partial \overline{\zeta }} & 
\frac{\text{$\partial $}^{2}F}{\partial \overline{\zeta }^{2}}
\end{array}
\right| \neq 0;  \label{zz}
\end{equation}

\noindent $\hat{D}=0$ when there is a critical point of the Lagrangian map
defined shortly.

The Lagrangian submanifold obtained from $F$ is defined in the following way:

First we have $p_{i}=\frac{\partial F}{\partial x^{i}}.$ (Note that this
involves only the explicit $x^{i}$ dependence in $F$ since any implicit
dependence, via the ($\zeta ,\bar{\zeta}),$ does not enter into the
definition of $p_{i}$ because of Eq.(\ref{s}).) Now depending on which case, 
$\#1,2$ or $3$ is relevant, we eliminate two of the five variables $%
(x^{i},\zeta ,\bar{\zeta}),$ in the $p_{i}=\frac{\partial F}{\partial x^{i}}%
(x^{i},\zeta ,\bar{\zeta}).$ This leaves the result that the six coordinates
of $T^{*}M,$ can be expressed in terms of three parameters, thus defining a
three dimensional submanifold in each of the cases;

\begin{itemize}
\item  In case 1, $p_{i}=$ $P_{i}(x^{i}),$ $\quad x^{i}=x^{i};$ the three
parameters \{$x^{i}\}=\chi ^{\alpha }.$

\item  In case 2, $p_{i}=$ $P_{i}(x^{A},\bar{\zeta}),$ $\quad x^{A}=x^{A},$ $%
\quad x^{J}=$ $X^{J}(x^{A},\bar{\zeta});$ the three parameters \{$x^{A},\bar{%
\zeta}\}=\chi ^{\alpha }.$

\item  In case 3, $p_{i}=$ $P_{i}(x^{J},\zeta ,\bar{\zeta}),$ $\quad x^{A}=$ 
$X^{A}(x^{J},\zeta ,\bar{\zeta}),\quad x^{J}=x^{J};$ the three parameters \{$%
x^{J},\zeta ,\bar{\zeta}\}=\chi ^{\alpha }.$
\end{itemize}

To simplify the discussion, we have referred to the three parameters, in
each of the cases, simply as $\chi ^{\alpha }.$ In each case we thus have

\begin{equation}
x^{i}=X^{i}(\chi ^{\alpha }),\text{\qquad }p_{i}=P_{i}(\chi ^{\alpha }).
\end{equation}

Of immediate relevance to us is the projection (Lagrange map) of the
Lagrange submanifold into the configuration space which becomes in each of
the cases, $x^{i}=X^{i}(\chi ^{\alpha })$ or

\begin{itemize}
\item  In case 1, $x^{i}=x^{i},$ trivial diffeomorphism

\item  In case 2, $x^{A}=x^{A},\quad $ $x^{J}=$ $X^{J}(x^{A},\bar{\zeta})$

\item  In case 3, $x^{A}=$ $X^{A}(x^{J},\zeta ,\bar{\zeta}),$ $\quad
x^{J}=x^{J}$
\end{itemize}

The caustics of this problem are the regions in the configuration space
where the mappings $\#1,2$ or $3$ have rank two or one; i.e., where the
Jacobian of the mapping vanishes. They occur when the determinant $\hat{D}%
=0. $ The inverse image to the caustics in the parameter space are referred
to as the critical points of the Lagrange map. It is clear that in case $1,$
the Jacobian is one and rank reduction can only occur in cases $2$ and $3.$

This treatment of caustics can be extended into the full four-space by
eliminating, in the expression for $t=T(f;x^{i},\zeta ,\bar{\zeta}),$ or in
the implicit version, $f=F(t,x^{i},\zeta ,\bar{\zeta}),$ two of the five
variables $(x^{i},\zeta ,\bar{\zeta})$ via the cases $\#1,2,3.$ This results
in $t$ now being a function of the three parameters, $\chi ^{\alpha }$.
Though we will not need the full theory here, this construction leads to a
seven dimensional manifold, $(t,x^{i},p_{i})$ \{an example of a contact
manifold\} and a three-dimensional submanifold of the contact manifold (a
Legendre submanifold) defined by

\begin{equation}
t=T(\chi ^{\alpha }),\text{\quad }x^{i}=X^{i}(\chi ^{\alpha }),\text{ \quad }%
p_{i}=P_{i}(\chi ^{\alpha })
\end{equation}
as well as the Legendre mapping from the Legendre submanifold, to
space-time, ($t,x^{i})$

\begin{equation}
\{t,x^{i},p_{i}\}(\chi ^{\alpha })\Rightarrow t=T(\chi ^{\alpha }),\quad
x^{i}=X^{i}(\chi ^{\alpha })  \label{u}
\end{equation}
a three-surface in space-time - the ``big wavefronts'' in Arnold 's language
- in our case a null surface. The singularities of the map Eq.(\ref{u}),
where the rank drops below three, are the null surface singularities. These
singularities, at fixed $t,$ are the (``small'') wavefront singularities of
the previous section.

We now return to the question of the determination of the function $%
f=F(t,x^{i},\zeta ,\bar{\zeta})$ of Eq.(\ref{r}) for use in the study of
null surfaces. Our choice will be, from Eq.(\ref{g}),

\begin{equation}
F(t,x^{i},\zeta ,\bar{\zeta})=S^{**}(t,x^{i},\zeta ,\overline{\zeta })\equiv
x^{a}\ell _{a}(\zeta ,\overline{\zeta })+\alpha (\zeta ,\overline{\zeta }).
\end{equation}

There are three independent reasons for this choice;

\begin{description}
\item  1. It was the method of generating an arbitrary null surface from the
complete solution, $x^{a}\ell _{a}(\zeta ,\overline{\zeta })$; see Sec. II..

\item  2. It was the method for the construction of a null surface such that
the generators were orthogonal to a given two-surface; see Sec. III..

\item  3. It arises from a variant of Fermat's Principle of stationary time:
Consider a timelike worldline, $\frak{L,}$ (in a Lorentzian space-time) of
the form, in local coordinates, $(x^{i}=$ constant, $t$ varies$)$ and a
two-dimensional space-like surface, $\frak{s}(\zeta ,\overline{\zeta }).$
Assume, locally, that from every point of $\frak{s}(\zeta ,\overline{\zeta })
$ there is a null geodesic that reaches $\frak{L}$ at a time $%
t=T(x^{i},\zeta ,\overline{\zeta }),$ then $t$ is extremized by those curves
that are normal to $\frak{s}(\zeta ,\overline{\zeta }).$ This result (which
will be described in detail elsewhere) follows from Schr\"{o}dinger's
derivation \cite{Sch} of the gravitational frequency shift.
\end{description}

Note that the rank condition on the matrix, Eq.(\ref{t}) is satisfied by
direct calculation.

\qquad From the discussion of generating families, we see that the treatment
of the null surfaces that we proposed, in Secs. II and III, namely to solve
for the $\zeta =\Upsilon (x,y,z),$ was really only valid for Case $1,$ but
we actually used a version of Case $3$ where the additional parameter $r$
was introduced in order not to single out any particular cartesian
coordinate. Case $1$ broke down precisely on the caustic, given by 
\begin{equation}
r^{2}=\text{\dh }^{2}\alpha \overline{\text{\dh }}^{2}\alpha
\end{equation}
which is where Cases $2$ and $3$ must be applied.

What follows is a straightforward application of Case 3 of these ideas - on
one patch. For completeness we repeat some of the earlier steps

\begin{quotation}
Starting with 
\begin{equation}
u=S^{**}(x^{a},\zeta ,\bar{\zeta})=x^{a}l_{a}(\zeta ,\bar{\zeta})+\alpha
(\zeta ,\bar{\zeta})  \label{v}
\end{equation}
\end{quotation}

\noindent \noindent then 
\begin{eqnarray}
\text{\dh }S^{**} &=&\frac{1}{\sqrt{2}(1+\zeta \bar{\zeta})}[\overline{\zeta 
}^{2}W-{\bar{W}}-2z\overline{\zeta }]+\text{\dh }\alpha ,  \nonumber \\
{\bar{\text{\dh }}}S^{**} &=&\frac{1}{\sqrt{2}(1+\zeta \bar{\zeta})}[\zeta
^{2}{\bar{W}}-W-2z\zeta ]+\overline{\text{\dh }}\alpha ,
\end{eqnarray}
with $W=(x+iy)$. From \dh $S^{**}=0$ and ${\bar{\text{\dh }}}S^{**}=0$, we
obtain, (from Case 3, where $x^{A}$ are $(x,y)$ or ($W$, $\overline{W}))$
that 
\begin{eqnarray}
W &=&\frac{\sqrt{2}({\bar{\text{\dh }}}\alpha +\zeta ^{2}\text{\dh }\alpha
)-2z\zeta }{1-\zeta \bar{\zeta}},  \label{w} \\
\bar{W} &=&\frac{\sqrt{2}(\text{\dh }\alpha +\bar{\zeta}^{2}{\bar{\text{ \dh 
}}}\alpha )-2z\bar{\zeta}}{1-\zeta \bar{\zeta}},  \nonumber
\end{eqnarray}

\noindent From Eq.(\ref{v}) we have that 
\begin{eqnarray}
p_{x} &=&\frac{\partial S^{**}}{\partial x}=-\frac{\zeta +\bar{\zeta}}{\sqrt{%
2}(1+\zeta \bar{\zeta})},  \nonumber \\
p_{y} &=&\frac{\partial S^{**}}{\partial y}=\frac{i(\zeta -\bar{\zeta})}{%
\sqrt{2}(1+\zeta \bar{\zeta})},  \nonumber \\
p_{z} &=&\frac{\partial S^{**}}{\partial z}=\frac{(1-\zeta \bar{\zeta})}{%
\sqrt{2}(1+\zeta \bar{\zeta})}
\end{eqnarray}
Taking $p=\sqrt{2}(p_{x}+ip_{y})$, we obtain

\begin{equation}
p=-\frac{2\zeta }{1+\zeta \bar{\zeta}}.
\end{equation}

The Lagrange submanifold, parametrized by $(z,\zeta ,\overline{\zeta }),$ is
give by

\begin{eqnarray}
z &=&z,  \label{x} \\
W &=&\frac{\sqrt{2}({\bar{\text{\dh }}}\alpha +\zeta ^{2}\text{\dh }\alpha
)-2z\zeta }{1-\zeta \bar{\zeta}},  \label{xx} \\
\bar{W} &=&\frac{\sqrt{2}(\text{\dh }\alpha +\bar{\zeta}^{2}{\bar{\text{ \dh 
}}}\alpha )-2z\bar{\zeta}}{1-\zeta \bar{\zeta}},  \label{3x} \\
p &=&-\frac{2\zeta }{1+\zeta \bar{\zeta}},  \label{xxxx} \\
\bar{p} &=&-\frac{2\overline{\zeta }}{1+\zeta \bar{\zeta}},  \label{xxxxx} \\
p_{z} &=&\frac{(1-\zeta \bar{\zeta})}{\sqrt{2}(1+\zeta \bar{\zeta})},
\label{xxxxxx}
\end{eqnarray}

The projection to the configuration space is given by 
\begin{eqnarray}
z &=&z,  \nonumber \\
W &=&\frac{\sqrt{2}({\bar{\text{\dh }}}\alpha +\zeta ^{2}\text{\dh }\alpha
)-2z\zeta }{1-\zeta \bar{\zeta}},  \nonumber \\
\bar{W} &=&\frac{\sqrt{2}(\text{\dh }\alpha +\bar{\zeta}^{2}{\bar{\text{ \dh 
}}}\alpha )-2z\bar{\zeta}}{1-\zeta \bar{\zeta}}.
\end{eqnarray}

The critical points of the Lagrange map are obtained from the condition that
the Jacobian 
\begin{equation}
J=\frac{\partial (z,W,\bar{W})}{\partial (z,\zeta ,\bar{\zeta})}
\end{equation}
vanishes. The vanishing of $J$ is equivalent to $D=r^{2}-$ \dh $^{2}\alpha $
\={\dh }$^{2}\alpha =0.$

To construct the Legendrian submanifold (in the seven dimensional contact
space, $(t,x^{i},p_{i}))$ we take the generating family $u=S^{**}(x^{a},%
\zeta ,\bar{\zeta})$ where $u$ is constant and solve for $t$ expressing the
contact coordinate $t$ in terms of the three parameters $(z,\zeta ,\bar{\zeta%
})$ by

\begin{equation}
t=\frac{\zeta \bar{W}+\bar{\zeta}W-z(1-\zeta \bar{\zeta})}{(1+\zeta \bar{%
\zeta})}+\sqrt{2}[u-\alpha (\zeta ,\bar{\zeta})]  \label{y}
\end{equation}
with

\begin{eqnarray}
W &=&\frac{\sqrt{2}({\bar{\text{\dh }}}\alpha +\zeta ^{2}\text{\dh }\alpha
)-2z\zeta }{1-\zeta \bar{\zeta}},  \nonumber \\
\bar{W} &=&\frac{\sqrt{2}(\text{\dh }\alpha +\bar{\zeta}^{2}{\bar{\text{ \dh 
}}}\alpha )-2z\bar{\zeta}}{1-\zeta \bar{\zeta}}.
\end{eqnarray}

The full Legendre submanifold is then given by Eqs.(\ref{x}), (\ref{xx}), (%
\ref{3x}), (\ref{xxxx}), (\ref{xxxxx}), (\ref{xxxxxx}) and (\ref{y}).

Note that this entire construction, using Case \#3, was valid where $1-\zeta 
\bar{\zeta}\neq 0$ (or equivalently where $p_{z}\neq 0).$ To include the
region where $p_{z}=0$, a different choice of parametrization would be
necessary, e.g. $(x$, $\zeta ,\bar{\zeta}),$ which is valid in the region
where $p_{x}\neq 0$ or $(y$, $\zeta ,\bar{\zeta}),$ valid where $p_{y}\neq 0$%
.

Using the example \#2, from Sec. IV, given by

\begin{equation}
S^{**}(x^{a},\zeta ,\bar{\zeta})=x^{a}l_{a}(\zeta ,\bar{\zeta})+\alpha
(\zeta ,\bar{\zeta}),\text{ }\alpha =-\frac{(1-\zeta \bar{\zeta})(\zeta +%
\bar{\zeta})}{(1+\zeta \bar{\zeta})^{2}}
\end{equation}
with 
\begin{equation}
\text{\dh }S^{**}=0\Leftrightarrow \{x(-1+\overline{\zeta }^{2})+iy(1+\bar{%
\zeta}^{2})-2z\bar{\zeta}](1+\zeta \bar{\zeta})=\sqrt{2}[1+\bar{\zeta}%
^{3}\zeta -3\bar{\zeta}(\zeta +\bar{\zeta})]  \label{yy}
\end{equation}
and 
\begin{equation}
\overline{\text{\dh }}S^{**}=0\Leftrightarrow \{x(-1+\zeta ^{2})-iy(1+\zeta
^{2})-2z\zeta ](1+\zeta \bar{\zeta})=\sqrt{2}[1+\zeta ^{3}\bar{\zeta}-3\zeta
(\zeta +\bar{\zeta})]  \label{yyy}
\end{equation}
one could try to solve for different pairs from the set $(x,y,z,\zeta ,\bar{%
\zeta})$. When $D\neq 0$ one could always solve for ($\zeta ,\bar{\zeta}),$
though in general there would be more than one solution; i.e., for fixed $%
(x,y,z)$ there would in general be more than one ray going thru that
space-point, either at the same or at different times. Alternately one could
try to solve in different regions for $(x,y),(y,z),(z,x),$ etc. Solving for $%
(x,y)$ or $W$ we have that the Lagrange map (from the ($z,\zeta ,\bar{\zeta}%
) $ parameter space) becomes

\begin{eqnarray}
W &=&\frac{\sqrt{2}(2\zeta ^{2}{+4}\zeta \bar{\zeta}-1-\zeta ^{2}\bar{\zeta}%
^{2})-2z\zeta (1+\zeta \bar{\zeta})}{1-\zeta ^{2}\bar{\zeta}^{2}},\quad \\
\overline{W} &=&\frac{\sqrt{2}(2\overline{\zeta }^{2}{+4}\zeta \bar{\zeta}%
-1-\zeta ^{2}\bar{\zeta}^{2})-2z\overline{\zeta }(1+\zeta \bar{\zeta})}{%
1-\zeta ^{2}\bar{\zeta}^{2}}
\end{eqnarray}
which in turn becomes the Legendre map when the contact coordinate $t$ is
added in;

\begin{equation}
t=\sqrt{2}\{u+\frac{1}{(1-\zeta \bar{\zeta})(1+\zeta \bar{\zeta})^{2}}[
4\zeta \bar{\zeta}(\zeta +\bar{\zeta})-\frac{z}{\sqrt{2}}(1+3\zeta \bar{\zeta%
}+3\zeta ^{2}\bar{\zeta}^{2}+\zeta ^{3}\bar{\zeta}^{3})]\}.
\end{equation}

Though none of the details of this analysis is particularly enlightening, it
nevertheless shows how in principle one constructs the Lagrange submanifold
and the Lagrange map even in the presence of the caustics.

\section{Families of Foliations}

In this section we will generalize, in the following sense, the ideas of
Section II. Recently there has been a reformulation of General Relativity,
referred to as the Null Surface Formulation (NSF) where the basic idea has
been to use a family (a spheres worth) of null foliations of space-time, so
that there are a spheres worth of null surfaces passing thru each point of
space-time.. These surfaces are described as the level surfaces of the
function

\qquad \qquad \qquad $u=Z(x^{a},\zeta ,\overline{\zeta })$;

\noindent $x^{a}$ are the local space-time coordinates and ($\zeta ,%
\overline{\zeta })$ are the complex stereographic coordinate on the sphere
which labels the family of foliations. The function Z, for every fixed value
of ($\zeta ,\overline{\zeta })$, satisfies the eikonal equation,\qquad
\qquad 
\begin{equation}
g^{ab}\partial _{a}Z\partial _{b}Z=0.  \label{10}
\end{equation}

Knowing these families of foliations one can construct the (conformal)
metric in terms of $Z$. The idea was then to express the Einstein equations
in terms of these surfaces, i.e., in terms of $Z$ and a conformal factor.
Though this was successfully accomplished, a technical difficulty in fully
understanding the equations arose due to the fact that the null surfaces
developed singularities (caustics) and self-intersections. It was clear that
the development of caustics was a generic feature of the equations but it
was not at all clear how to see and study their existence directly in terms
of the function $Z(x^{a},\zeta ,\overline{\zeta })$ and its derivatives. In
this section, we will study, in flat space, the construction of such
families and show explicitly how to calculate the structure of the null
surface singularities (the caustics and wave-front singularities) directly
in terms of the Z function.

\textit{Locally} (up to first derivatives) there is no direct curvature
involvement in the eikonal equation, so that the form of the caustics in
terms of $Z$ should apply equally in curved space as in Minkowski space. The
results obtained here for Minkowski space will thus likely apply to the
curved space situation.

Starting with the two-parameter family of plane waves used earlier, $%
Z_{0}(x^{a},\zeta ,\overline{\zeta })=x^{a}\ell _{a}(\zeta ,\overline{\zeta }%
),$ we will first construct a general two-parameter family of solutions to
the flat-space eikonal equation, $Z(x^{a},\mu ,\overline{\mu }),$ with $(\mu
,\overline{\mu })$ parametrizing the sphere; we then study the singularities
and caustics of this new family.

We begin by generalizing Eq.(\ref{g}), namely

\qquad \qquad \qquad $S^{**}=x^{a}\ell _{a}(\zeta ,\overline{\zeta }
)+\alpha (\zeta ,\overline{\zeta })$

\noindent by writing $\alpha $ as a function on $S^{2}$x$S^{2}$; i.e., as $%
\alpha $ = $\alpha $($\zeta ,\overline{\zeta }$, $\mu ,\overline{\mu }$) and
then repeating the earlier procedure of setting to zero, the \dh\ and $%
\overline{\text{\dh }}$ derivatives with respect to the ($\zeta ,\overline{%
\zeta })$; i.e., considering

\begin{equation}
u=Z^{**}(x,\zeta ,\overline{\zeta },\mu ,\overline{\mu })=x^{a}\ell
_{a}(\zeta ,\overline{\zeta })+\alpha (\zeta ,\overline{\zeta },\mu ,%
\overline{\mu })  \label{11}
\end{equation}
and

\begin{equation}
\text{\dh }_{\zeta }Z^{**}=\overline{\text{\dh }}_{\zeta }Z^{**}=0
\label{11a}
\end{equation}
and then solving them (when possible) for $(\zeta ,\overline{\zeta })$
obtaining$\qquad \qquad $%
\begin{equation}
\zeta =\Upsilon (x,y,z,\mu ,\overline{\mu }),\qquad \overline{\zeta }=%
\overline{\Upsilon }(x,y,z,\mu ,\overline{\mu })  \label{11b}
\end{equation}
so that when substituted into Eq.(\ref{11}) we obtain the new family which
depends on the choice of $\alpha (\zeta ,\overline{\zeta },\mu ,\overline{%
\mu });$

\begin{equation}
Z(x^{a},\mu ,\overline{\mu })=x^{a}l_{a}(\Upsilon ,\overline{\Upsilon }%
)+\alpha (\Upsilon ,\overline{\Upsilon },\mu ,\overline{\mu }).  \label{15}
\end{equation}

(Alternatively we could use the different cases of Sec. V, when one can not
solve for ($\zeta ,\overline{\zeta })$.) It is obvious from the previous
discussion that Eq.(\ref{15}) satisfies the eikonal equation for each fixed
value of ($\mu ,\overline{\mu )}$. All we have done so far is create a new
spheres worth of null foliations (wavefront families) of Minkowski space -
different from the plane wave case of $S=x^{a}\ell _{a}(\zeta ,\overline{%
\zeta }).$ As in the earlier sections we could have analyzed the null
surfaces for each value of ($\mu ,\overline{\mu })$ separately but now in
this generalization the null surfaces are smoothly connected to each other
through the variable ($\mu ,\overline{\mu })$ and it becomes of interest to
see the development of the caustics via the variation of the ($\mu ,%
\overline{\mu })$, or through the ($\mu ,\overline{\mu })$ derivatives.

\begin{remark}
We will use, respectively, the notation (\dh $_{\mu },\overline{\text{\dh }}%
_{\mu })$ for the eth and ethbar derivatives with respect to the variables ($%
\mu ,\overline{\mu })$ and (\dh $_{\zeta },\overline{\text{\dh }}_{\zeta })$
for the variables ($\zeta ,\overline{\zeta }).$
\end{remark}

We begin by defining several derivatives of $Z$;$\qquad \qquad $%
\begin{equation}
\omega =\text{\dh }_{\mu }Z,\text{ }\overline{\omega }=\overline{\text{\dh }}
_{\mu }Z
\end{equation}
$\qquad \qquad \qquad \qquad \qquad \qquad \qquad $%
\begin{equation}
R=\overline{\text{\dh }}_{\mu }\text{\dh }_{\mu }Z
\end{equation}

\begin{equation}
\Lambda =\text{\dh }_{\mu }^{2}Z,\text{ }\overline{\Lambda }=\overline{\text{%
\dh }}_{\mu }^{2}Z.
\end{equation}

A level surface of Z, with fixed ($\mu ,\overline{\mu })$, is ruled by null
geodesics, whose tangent vectors are given by $\ell _{a}(\Upsilon ,\overline{
\Upsilon }$); A pencil of rays defined from a pair of geodesic deviation
vectors (from a given geodesic) has an area $A$ that can be given\cite{KN2}
up to a proportionality by

$\qquad \qquad \qquad \qquad $%
\begin{equation}
A=K\frac{\Omega ^{-2}}{\sqrt{(1-\Lambda _{,1}\overline{\Lambda }_{,1})}}
\label{16}
\end{equation}
where $K$ is a constant determined by the initial area and

$\qquad \qquad \qquad \Omega ^{2}$ = $\ell ^{a}R,_{a}$

\noindent and

$\qquad \qquad \qquad \Lambda _{,1}$= $\Omega ^{-2}$ $\ell ^{a}\Lambda $,$%
_{a}$.

The derivation of Eq.(\ref{16}) is lengthy and will not be given here. It
will however be shown, in this case to be proportional to the area.

We now want to see the behavior of $\omega ,$ $\Lambda $ and $R,$ as well as
the area $A,$ in the neighborhood of a caustic.

By direct calculation we have, from Eqs.(\ref{15} and \ref{11a}), that$%
\qquad \qquad $%
\begin{equation}
\omega =\text{\dh }_{\mu }\alpha ,\text{ }\overline{\omega }=\overline{\text{%
\dh }}_{\mu }\alpha ,  \label{17}
\end{equation}

\noindent and hence is singularity free.

After a rather lengthy calculation, using Eqs.(\ref{11}) and (\ref{11a}), we
obtain

\begin{eqnarray}
R &=&\text{\dh }_{\mu }\overline{\text{\dh }}_{\mu }\alpha +\frac{1}{D}\{(%
\text{\dh }_{\zeta }\text{\dh }_{\mu }\alpha )[(\text{\dh }_{\zeta }%
\overline{\text{\dh }}_{\mu }\alpha )(\overline{\text{\dh }}_{\zeta }^{2}u)-(%
\overline{\text{\dh }}_{\zeta }\overline{\text{\dh }}_{\mu }\alpha )(\text{%
\dh }_{\zeta }\overline{\text{\dh }}_{\zeta }u)]  \label{20} \\
&&\text{ }+(\overline{\text{\dh }}_{\zeta }\text{\dh }_{\mu }\alpha )[(%
\overline{\text{\dh }}_{\zeta }\overline{\text{\dh }}_{\mu }\alpha )(\text{%
\dh }_{\zeta }^{2}u)-(\text{\dh }_{\zeta }\overline{\text{\dh }}_{\mu
}\alpha )(\text{\dh }_{\zeta }\overline{\text{\dh }}_{\zeta }u)]\}. 
\nonumber
\end{eqnarray}
where

\begin{equation}
D=(\text{\dh }_{\zeta }\overline{\text{\dh }}_{\zeta }u)^{2}-\text{\dh }
_{\zeta }^{2}u\overline{\text{\dh }}_{\zeta }^{2}u.  \label{19}
\end{equation}

In a similar way we obtain$\qquad $%
\begin{eqnarray}
\Lambda &=&\text{\dh }_{\mu }^{2}\alpha +\frac{1}{D}\{(\text{\dh }_{\zeta }%
\text{\dh }_{\mu }\alpha )[(\text{\dh }_{\zeta }\text{\dh }_{\mu }\alpha )(%
\overline{\text{\dh }}_{\zeta }^{2}u)-(\overline{\text{\dh }}_{\zeta }\text{%
\dh }_{\mu }\alpha )(\text{\dh }_{\zeta }\overline{\text{\dh }}_{\zeta }u)]
\label{21} \\
&&\text{ }+(\overline{\text{\dh }}_{\zeta }\text{\dh }_{\mu }\alpha )[(%
\overline{\text{\dh }}_{\zeta }\text{\dh }_{\mu }\alpha )(\text{\dh }_{\zeta
}^{2}u)-(\text{\dh }_{\zeta }\text{\dh }_{\mu }\alpha )(\text{\dh }_{\zeta }%
\overline{\text{\dh }}_{\zeta }u)]\}.  \nonumber
\end{eqnarray}

First we see that for fixed values of ($\mu ,\overline{\mu })$, $D$ from Eq.(%
\ref{19}), is the same as in Eq.(\ref{oo}), namely $D=r^{2}-$ \dh $_{\zeta
}^{2}\alpha \overline{\text{\dh }}_{\zeta }^{2}\alpha $ and hence vanishes
at the caustic. We can now see that $\omega $ is regular at the caustic
while both $R$ and $\Lambda $ have singularities of the form $D^{-1}$ at the
caustic.

In order to find the area $A$ we first need $R,_{a}$ and $\Lambda ,_{a}.$
After a lengthy calculation we obtain

\begin{eqnarray}
\Omega ^{2} &\equiv &\ell ^{a}R,_{a}=\frac{1}{D^{2}}\{[(\text{\dh }_{\zeta
}^{2}u)(\overline{\text{\dh }}_{\zeta }^{2}u)+(\text{\dh }_{\zeta }\overline{%
\text{\dh }}_{\zeta }u)^{2}][(\text{\dh }_{\zeta }\text{\dh }_{\mu }\alpha )(%
\overline{\text{\dh }}_{\zeta }\overline{\text{\dh }}_{\mu }\alpha )+(%
\overline{\text{\dh }}_{\zeta }\text{\dh }_{\mu }\alpha )(\text{\dh }_{\zeta
}\overline{\text{\dh }}_{\mu }\alpha )]  \nonumber \\
&&-2[(\text{\dh }_{\zeta }\text{\dh }_{\mu }\alpha )(\text{\dh }_{\zeta }%
\overline{\text{\dh }}_{\mu }\alpha )(\overline{\text{\dh }}_{\zeta }^{2}u)+(%
\overline{\text{\dh }}_{\zeta }\overline{\text{\dh }}_{\mu }\alpha )(%
\overline{\text{\dh }}_{\zeta }\text{\dh }_{\mu }\alpha )(\text{\dh }_{\zeta
}^{2}u)](\text{\dh }_{\zeta }\overline{\text{\dh }}_{\zeta }u)\}  \label{22}
\end{eqnarray}

\noindent and$\qquad $%
\begin{eqnarray}
\Omega ^{2}\Lambda _{,1} &\equiv &\ell ^{a}\Lambda ,_{a}=\frac{2}{D^{2}}\{[(%
\text{\dh }_{\zeta }^{2}u)(\overline{\text{\dh }}_{\zeta }^{2}u)+(\text{\dh }%
_{\zeta }\overline{\text{\dh }}_{\zeta }u)^{2}](\text{\dh }_{\zeta }\text{%
\dh }_{\mu }\alpha )(\overline{\text{\dh }}_{\zeta }\text{\dh }_{\mu }\alpha
)  \nonumber \\
&&-[(\text{\dh }_{\zeta }\text{\dh }_{\mu }\alpha )^{2}(\overline{\text{\dh }%
}_{\zeta }^{2}u)+(\overline{\text{\dh }}_{\zeta }\text{\dh }_{\mu }\alpha
)^{2}(\text{\dh }_{\zeta }^{2}u)](\text{\dh }_{\zeta }\overline{\text{\dh }}%
_{\zeta }u)\}  \label{23}
\end{eqnarray}
Though it is not immediately obvious, from Eqs.(\ref{22}) and (\ref{23}) one
can show that ($1-\Lambda _{,1}\overline{\Lambda }_{,1})$ is proportional to 
$D^{2}$ or that \TEXTsymbol{\vert}$\Lambda _{,1}|$ $\Rightarrow 1+O(D)$ at
the caustic. From these results we have that\qquad 
\begin{equation}
A=K\frac{\Omega ^{-2}}{\sqrt{(1-\Lambda _{,1}\overline{\Lambda }_{,1})}}=%
\frac{\pm KD}{\hat{K}}  \label{24}
\end{equation}
with

\qquad \qquad $\qquad \qquad \qquad \qquad \hat{K}=|$\dh $_{\zeta }$\dh $%
_{\mu }\alpha |^{2}-|\overline{\text{\dh }}_{\zeta }$\dh $_{\mu }\alpha
|^{2}.$

From Eqs. (\ref{11a}) and (\ref{11b}), we have that $\hat{K}=\hat{K}%
(\Upsilon ,\overline{\Upsilon })$ from which it can be shown that

$\qquad \qquad \qquad \qquad \qquad \ell ^{a}\hat{K},_{a}=0,$

\noindent i.e., $\hat{K}$ is constant along the geodesic flow. If we chose $%
K=\pm \hat{K}$ we have that $A=D,$ in agreement with the area obtained in
section III from geodesic deviation.

Several important observations can now be made:

1. In our particular case of flat space, we have seen that the quantities $R$
and $\Lambda $ diverge as $D^{-1}$ at the caustic. It appears virtually
certain that this is a general result and remains true in a general curved
space.

2. As was to be expected the Area of a pencil of null geodesics vanishes at
the caustic. This is clearly true in general and the result here is a
confirmation that the Eq.(\ref{16}) really is the area formula.

3. The quantity $\Omega $ (which plays a central role in the NSF version of
GR) diverges as $D^{-1}$ at the caustic.

4. Though $\Lambda $ diverges at the caustic, the absolute value of its
weighted derivative

\qquad \qquad \TEXTsymbol{\vert}$\Lambda ,_{1}$\TEXTsymbol{\vert} = 
\TEXTsymbol{\vert}$\Omega ^{-2}\Lambda ,_{a}$ $\ell ^{a}$\TEXTsymbol{\vert}

\noindent approaches one as $1-O(D)$. From this one sees that $\Lambda
,_{a}\ell ^{a}$ diverges as $D^{-2}$.

\section{Discussion}

Our main interest in the study of wavefronts and their associated null
surfaces, lies in our desire to understand and describe their singularity
structure in curved Lorentzian manifolds and in particular to find the most
appropriate variables and representations for their analysis. Though locally
the classification of generic singularities and caustics is complete and is
the same in both flat and curved spaces\cite{HF}, however, in general
spaces, curvature effects are large and must eventually be taken into
account for global questions. (For example, the structure of the lightcone
in a curved space is very different from that of a lightcone in flat space.)
The present work is intended to begin this study with the description of
singular, global, asymptotically spherical, null surfaces in flat spaces. A
follow-up second paper, will be devoted to \textit{the same }issues as here
but in asymptotically flat space-times. We will see that beginning with a
two-parameter family of solutions of the eikonal equation - analogous to the
plane wave solutions of flat space-time - it will be possible to construct
any other null surface and then analyze its singularity structure. In
particular, it is possible to construct, in terms of the two parameter
family, the light cone of any space-time point. These insights are important
for applications of the null surface formulation of GR \cite{NSF1,NSF2}.

\section{Acknowledgment}

We would like to thank Carlos Kozameh for many very useful arguments and
even for some peaceful conversations. Also we want to point out that there
has been a fair amount of unwitting and uncoordinated overlap between the
work of the present authors (in Secs. II and III) and the group of Bishop,
Gomez, Lehner, Szilagyi and Winicour\cite{Wetal}. Since the motivation,
notation and final results of the work of the two groups are quite
different, it has been decided to keep the work separate and publish
independently.

We also thank the referee, Juergen Ehlers, for valuable improvements and
corrections.

SF and EN thank the NSF for support under grant \#PHY 92-05109. GSO
acknowledges the financial support from the Sistema Nacional de
Investigadores and from the Consejo Nacional de Ciencia y Tecnologia
(CONACyT) Mexico.

\end{document}